\input{aipcheck}

\documentclass[
    ,final            
  ]
  {aipproc}

\layoutstyle{8x11double}

\newcommand{\be}{\begin{equation}}
\newcommand{\ee}{\end{equation}}

\newcommand{\eqs}[2]{Eqs. (\ref{#1}) \& (\ref{#2})}

\newcommand{\Fig}[1]{Fig. (\ref{#1})} 
 \newcommand{\eqa}{\begin{eqnarray}}
\newcommand{\eeq}{\end{eqnarray}}  
\newcommand{\eqsto}[2]{Eqs. (\ref{#1}) to (\ref{#2})}

\begin{document}

\title{Self-consistent Simulations of Plasma-Neutral in a Partially Ionized Astrophysical Turbulent Plasma}



\author{Dastgeer Shaikh\footnote{\tt Email:dastgeer.shaikh@uah.edu}}{
  address={Department of Physics and Center for Space Plasma and Aeronomy Research (CSPAR), \\
The University of Alabama in Huntsville, Huntsville, AL 35899, USA}
}

\author{G. P. Zank}{}

\begin{abstract}
A local turbulence model is developed to study energy cascades in the
heliosheath and outer heliosphere (OH) based on self-consistent
two-dimensional fluid simulations. The model describes a partially
ionized magnetofluid OH that couples a neutral hydrogen fluid with a
plasma primarily through charge-exchange interactions. Charge-exchange
interactions are ubiquitous in warm heliospheric plasma, and the
strength of the interaction depends largely on the relative speed
between the plasma and the neutral fluid. Unlike small-length scale
linear collisional dissipation in a single fluid, charge-exchange
processes introduce channels that can be effective on a variety of
length scales that depend on the neutral and plasma densities,
temperature, relative velocities, charge-exchange cross section, and
the characteristic length scales. We find, from scaling arguments and
nonlinear coupled fluid simulations, that charge-exchange interactions
modify spectral transfer associated with large-scale energy-containing
eddies. Consequently, the turbulent cascade rate prolongs spectral
transfer among inertial range turbulent modes. Turbulent spectra
associated with the neutral and plasma fluids are therefore steeper
than those predicted by Kolmogorov's phenomenology. Our work is
important in the context of the global heliospheric interaction, the
energization and transport of cosmic rays, gamma-ray bursts,
interstellar density spectra, etc.  Furthermore, the plasma-neutral
coupling is crucial in understanding the energy dissipation mechanism
in molecular clouds and star formation processes.
\end{abstract}

\maketitle


\section{1. Introduction}
In many circumstances, astrophysical and helioshperic plasmas are
characterized by partially ionized gases in which magnetized plasma of
protons/electrons and neutral particles co-exist. Number densities of
these components and interaction processes can vary. The plasma and
the neutral particles especially in the heliospheric plasma interact
mutually through charge exchange, i.e., a proton sufficiently near a
neutral hydrogen atom can capture its electron, thereby creating a new
neutral and a proton. This process conserves plasma and neutral
densities, but not momentum and energy.  The charge exchange mean free
path in the local heliospheric plasma is typically about the order of
50 AU \cite{zank1999}. Length-scales smaller than the charge exchange
mean free path can exhibit turbulent motion. The physics of
small-scale turbulent motions is complex and is important to e.g.
heating, the transport of cosmic rays, Fermi acceleration, the density
spectrum. However no self-consistent simulation models attempt to
directly address the multi-component and multiple-scale character of
the outer heliosphere and its associated turbulence. Here we present a
self-consistent plasma-neutral turbulence simulation model based on
Fourier spectral techniques that are widely employed in the studies of
neutral and plasma fluids. We concentrate on some of the most
fundamental aspects of partially ionized heliospheric plasma
turbulence. In section 2, we discuss the equations of the coupled
plasma-neutral fluid. Section 3 describes results of nonlinear
self-consistent fluid simulations that evolve plasma and neutral
fluids in the presence of charge exchange forces. The charge exchange
spectra are discussed in section 4.  Conclusions are presented in
section 5.

\section{2. Plasma Neutral Coupled Fluid Model}
Our model simulates the partially ionized plasma in a local
region. The fluid model describing nonlinear turbulent processes, in
the presence of charge exchange, can be cast into plasma density
($\rho_p$), velocity (${\bf U}_p$), magnetic field (${\bf B}$),
pressure ($P_p$) components according to the conservative form
\be
\label{mhd}
 \frac{\partial {\bf F}_p}{\partial t} + \nabla \cdot {\bf Q}_p={\cal Q}_{p,n},
\ee
where,
\[{\bf F}_p=
\left[ 
\begin{array}{c}
\rho_p  \\
\rho_p {\bf U}_p  \\
{\bf B} \\
e_p
  \end{array}
\right], 
{\bf Q}_p=
\left[ 
\begin{array}{c}
\rho_p {\bf U}_p  \\
\rho_p {\bf U}_p {\bf U}_p+ \frac{P_p}{\gamma-1}+\frac{B^2}{8\pi}-{\bf B}{\bf B} \\
{\bf U}_p{\bf B} -{\bf B}{\bf U}_p\\
e_p{\bf U}_p
-{\bf B}({\bf U}_p \cdot {\bf B})
  \end{array}
\right], \]
\[{\cal Q}_{p,n}=
\left[ 
\begin{array}{c}
0  \\
{\bf Q}_M({\bf U}_p,{\bf V}_n, \rho_p, \rho_n, T_n, T_p)   \\
0 \\
Q_E({\bf U}_p,{\bf V}_n,\rho_p, \rho_n, T_n, T_p)
  \end{array}
\right]
\] 
and
\[ e_p=\frac{1}{2}\rho_p U_p^2 + \frac{P_p}{\gamma-1}+\frac{B^2}{8\pi}.\]
The above set of plasma equations is supplemented by $\nabla \cdot {\bf
B}=0$ and is coupled self-consistently to the  neutral density
($\rho_n$), velocity (${\bf V}_n$) and pressure ($P_n$) through a set
of hydrodynamic fluid equations,
\be
\label{hd}
 \frac{\partial {\bf F}_n}{\partial t} + \nabla \cdot {\bf Q}_n={\cal Q}_{n,p},
\ee
where,
\[{\bf F}_n=
\left[ 
\begin{array}{c}
\rho_n  \\
\rho_n {\bf V}_n  \\
e_n
  \end{array}
\right], 
{\bf Q}_n=
\left[ 
\begin{array}{c}
\rho_n {\bf V}_n  \\
\rho_n {\bf V}_n {\bf V}_n+ \frac{P_n}{\gamma-1} \\
e_n{\bf V}_n
  \end{array}
\right],\]
\[{\cal Q}_{n,p}=
\left[ 
\begin{array}{c}
0  \\
{\bf Q}_M({\bf V}_n,{\bf U}_p, \rho_p, \rho_n, T_n, T_p)   \\
Q_E({\bf V}_n,{\bf U}_p,\rho_p, \rho_n, T_n, T_p)
  \end{array}
\right],
\] 
\[e_n= \frac{1}{2}\rho_n V_n^2 + \frac{P_n}{\gamma-1}.\]
Equations (\ref{mhd}) to (\ref{hd}) form an entirely self-consistent
description of the coupled plasma-neutral turbulent fluid.  The
charge-exchange momentum sources in the plasma and the neutral fluids,
i.e. Eqs. (\ref{mhd}) and (\ref{hd}), are described respectively by
terms ${\bf Q}_M({\bf U}_p,{\bf V}_n,\rho_p, \rho_n, T_n, T_p)$ and
${\bf Q}_M({\bf V}_n,{\bf U}_p,\rho_p, \rho_n, T_n, T_p)$. A swapping
of the plasma and the neutral fluid velocities in this representation
corresponds, for instance, to momentum changes (i.e. gain or loss) in
the plasma fluid as a result of charge exchange with the ISM neutral
atoms (i.e. ${\bf Q}_M({\bf U}_p,{\bf V}_n,\rho_p, \rho_n, T_n, T_p)$
in Eq. (\ref{mhd})). Similarly, momentum change in the neutral fluid
by virtue of charge exchange with the plasma ions is indicated by
${\bf Q}_M({\bf V}_n,{\bf U}_p,\rho_p, \rho_n, T_n, T_p)$ in
Eq. (\ref{hd}). For a complete description of charge exchange forces,
the readers can refer to our work in Ref. \cite{shaikh08}.

\subsection{3. Simulations}

A two-dimensional (2D) nonlinear fluid code was developed to
numerically integrate \eqsto{mhd}{hd}.  The spatial discretization in
our code uses a discrete Fourier representation of turbulent
fluctuations based on a pseudospectral method \cite{scheme}, while we
use a Runge Kutta 4 method for the temporal integration. All the
fluctuations are initialized isotropically (no mean fields are
assumed) with random phases and amplitudes in Fourier space.  This
algorithm ensures conservation of total energy and mean fluid density
per unit time in the absence of charge exchange and external random
forcing. Additionally, $\nabla
\cdot {\bf B}=0$ is satisfied at each time step.  Our code is
massively parallelized using Message Passing Interface (MPI) libraries
to facilitate higher resolution.  The 2D simulations are not only
computationally simpler and less expensive (compared with the full
3D), they offer significantly higher resolutions even on moderate size
small cluster machines like Beowulf. Since we are interested in
investgating the inertial range spectra, we restrict our simulations
to 2D. Our simulations however retain all three components of the
magnetic field and the background field is assumed along the
y-direction. This enables us to treat the magnetic field appropriately
The initial isotropic turbulent spectrum of fluctuations is chosen to
be close to $k^{-2}$ (where $k$ is characteristic mode) with random
phases in both $x$ and $y$ directions.  The choice of such (or even a
flatter than -2) spectrum does not influence the dynamical evolution
as the final state in our simulations progresses towards fully
developed turbulence.  While the turbulence code is evolved with time
steps resolved self-consistently by the coupled fluid motions, the
nonlinear interaction time scales associated with the plasma $1/({\bf
k} \cdot {\bf U}_p({\bf k}))$ and the neutral $1/({\bf k} \cdot {\bf
V}_n({\bf k}))$ fluids can obviously be disparate. Accordingly,
turbulent transport of energy in the plasma and the neutral ISM fluids
takes place on distinctively separate time scales.

\begin{figure}
\label{fig1}
  \includegraphics[height=.3\textheight]{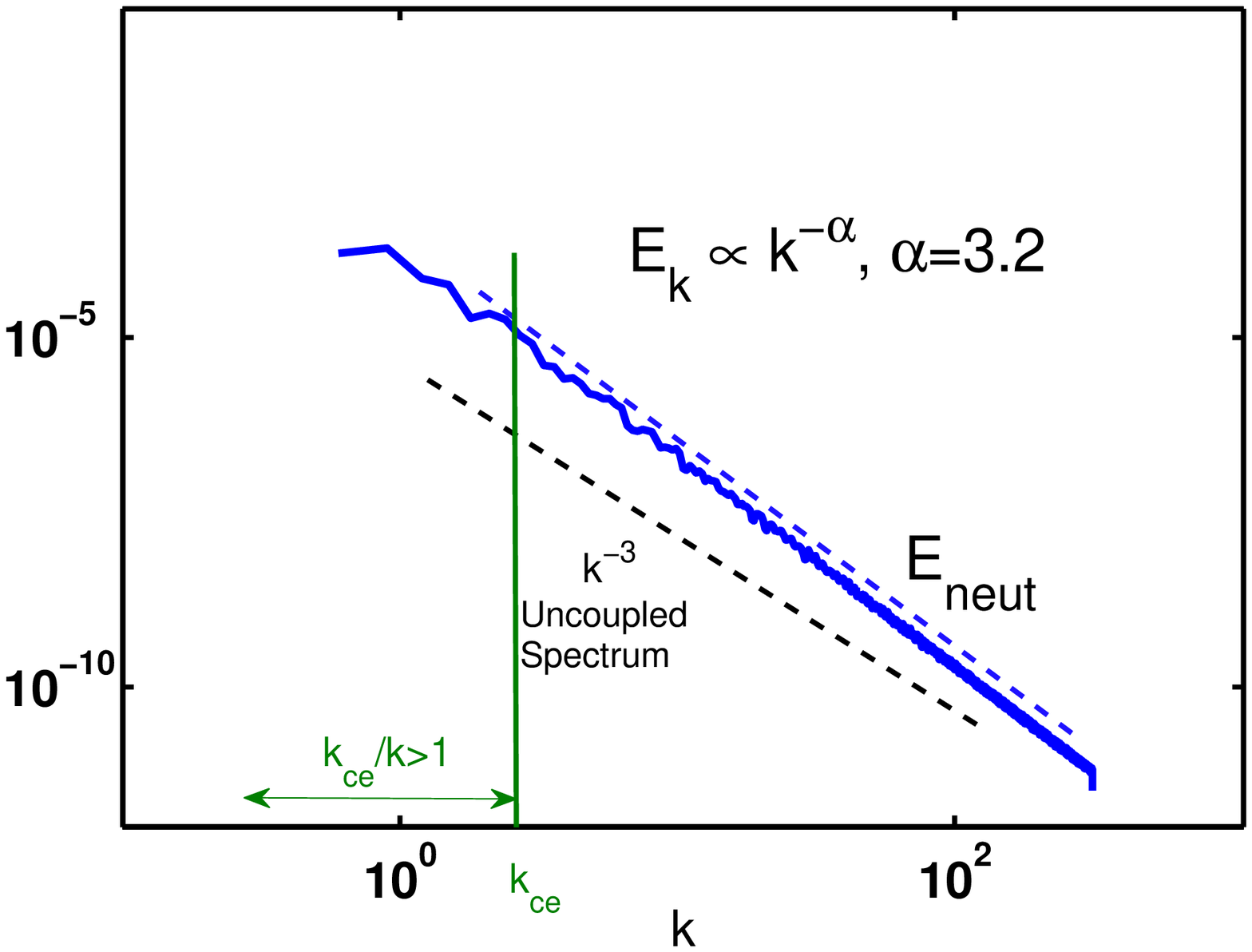}
\includegraphics[height=.3\textheight]{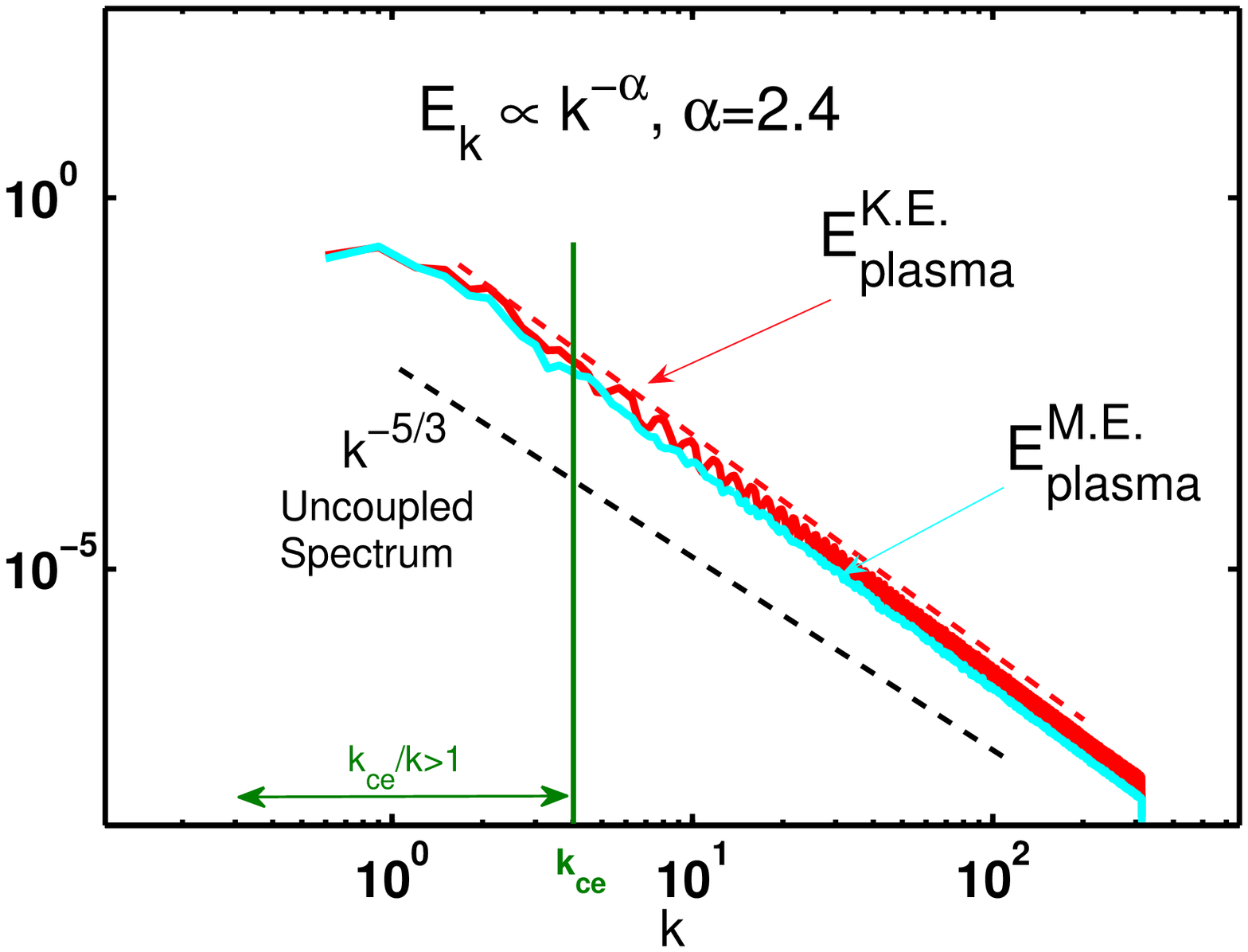}
  \caption{(LEFT) Spectra of neutral fluid. (RIGHT) Spectra of plasma fluid.}
\end{figure}

Spectral transfer in partially ionized fluid turbulence progresses
under the action of nonlinear interactions as well as charge exchange
sources. Energy cascades amongst turbulent eddies of various scale
sizes and between the plasma and the neutral fluids. In a freely
decaying case, plasma and neutral fluids evolve under the influence of
charge exchange forces which dramatically affect the energy cascades
in the inertial range. This is evident from \Fig{fig1} where
Kolmogorov-like \cite{kol,iros,krai} fully developed turbulent spectra
in the inertial range are shown respectively for the coupled neutral
and plasma fluids. The neutral fluid energy spectrum in the inertial
regime for the coupled system exhibits a $k^{-\alpha}$ spectrum, where
the spectral index $\alpha
\approx 3.2$. On the other hand, the spectral index for plasma
magnetic and kinetic energy spectra for the coupled system is $\alpha
\approx 2.4$.  The inertial range spectral indices for the coupled
plasma-neutral system in our simulations show a {\it significant}
deviation from their corresponding uncoupled analogues which are
respectively $-3$ and $-1.7$ for the neutral and plasma fluids.  The
spectral indices observed in our simulations can be followed from
Kolmogorov-like phenomenology \cite{kol,iros,krai}, as described in
the following.

The typical nonlinear interaction time-scale ($\tau_{nl}$) in ordinary
(i.e. uncoupled) plasma and/or neutral turbulence is given by
\be
\tau_{nl} \sim \frac{\ell_0}{v_\ell} \sim (kv_k)^{-1},
\ee 
where $v_k$ or $v_\ell$ is the velocity of turbulent eddies. In the
presence of charge exchange interactions, the ordinary nonlinear
interaction time-scale of fluid turbulence is modified by a factor
$k_{ce}/k$ such that the new nonlinear interaction time-scale ($\tau_{NL}$) in the partially
ionized heliospheric 
turbulence is now 
\be
\tau_{NL} \sim \frac{k_{ce}}{k} \frac{1}{kv_k}.
\ee 
On using the fact that $k_{ce}$ is typically larger than $k$ (or
$k_c$, the characteristic turbulent mode, as defined elsewhere in the
paper), i.e. $k_{ce}/k > 1$ in the heliosphere 
\citep{Florinski2003,Florinski2005,zank1999}, the new nonlinear time
is $k_{ce}/k$ times bigger than the old nonlinear time i.e.
$\tau_{NL} \sim (k_{ce}/k) \tau_{nl}$. This enhanced nonlinear
interaction time in the partially ionized plasma is likely to prolong
turbulent energy cascade rates.  It is because of this enhanced or
prolonged interaction time that a relatively large spectral transfer
of turbulent modes tends to steepen the inertial range turbulent
spectra in both plasma and neutral fluids. By extending the above
phenomenological analysis, one can deduce exact (analytic) spectral
indices of the inertial range decaying turbulent spectra, as
follows. The new nonlinear interaction time-scale of coupled
plasma-neutral turbulence can be rearranged as
\be
\tau_{NL} \sim
\frac{k_{ce}v_k}{kv_k}\frac{1}{kv_k}\sim \frac{\tau_{nl}^2}{\tau_{ce}},\ee 
where $\tau_{ce}\sim (k_{ce} v_k)^{-1}$ represents the charge exchange
time scale. The energy dissipation rate ($\varepsilon$) associated with the coupled
plasma-neutral system can be determined from $\varepsilon \sim
E_k/\tau_{NL}$, which leads to 
\be
\varepsilon \sim \frac{v_k^2}{k_{ce}/k^2 v_k}\sim \frac{k^2 v_k^3}{k_{ce}},
\ee
where $E_k$ is turbulent energy per unit mode.
According to the Kolmogorov theory, the spectral cascades are local in
$k$-space and the inertial range energy spectrum depends upon the
energy dissipation rates and the characteristic turbulent modes, such
that $E_k \sim \varepsilon^{\gamma} k^{\beta}$.  Upon substitution of
the above quantities and equating the power of identical bases, one
obtains
\be
E_k \sim \varepsilon^{2/3} k^{-7/3}\ee for the plasma spectrum (the
forward cascade inertial range). The spectral index associated with
this spectrum, i.e. $7/3 \approx 2.33$, is consistent with the plasma
spectrum observed in the (coupled plasma-neutral) simulations
(see \Fig{fig1}). Similar arguments in the context of neutral fluids,
when coupled with the plasma fluid in the heliosphere, lead to the energy
dissipation rates \be
\varepsilon \sim \frac{k^2 v_k^2}{k_{ce}/k^2 v_k}.\ee
 This further yields
 the forward cascade (neutral) energy spectrum 
\be
E_k \sim \varepsilon^{2/3} k^{-11/3}\ee
 which is close to the simulation result  shown in \Fig{fig1}.

\section{4. Charge exchange processes}

\begin{figure}[t]
\label{fig2}
  \includegraphics[height=.27\textheight]{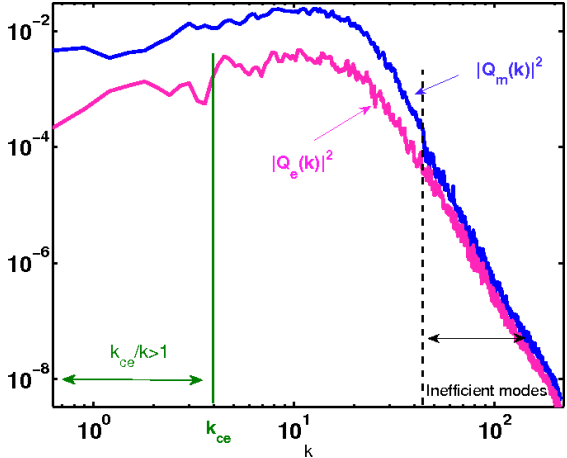}
  \caption{Charge exchange spectra}
\end{figure}

A key issue finally is to understand what role charge exchange modes
play in coupled plasma-neutral heliospheric turbulence as they result
essentially from the nonlinear charge exchange interactions. To
address this issue, we plot charge exchange sources associated with
the momentum and energy equations, \eqs{mhd}{hd}, in \Fig{fig2}. It
appears from \Fig{fig2} that spectral energy is transferred
predominantly at the larger scales by means of charge exchange mode
coupling processes. The latter couples the large-scales, or smaller
than $k_{ce}$ modes, efficiently to low-$k$ turbulent modes.  It is
primarily because of this $k_{ce}-k$ mode-coupling in the smaller $k$
part of the spectrum of the coupled plasma-neutral turbulence, that
energy is pumped efficiently at the lower $k$ inertial range turbulent
modes. The efficient coupling of the Fourier modes at low $k$'s
further enhances the nonlinear eddy time-scales associated with the
coupled plasma-neutral turbulence system which is consistent with the
scaling $\tau_{NL} \sim (k_{ce}/k) \tau_{nl}$, where $k_{ce}/k >1$, as
described above.  This consequently leads to the steepening of the
inertial range spectra observed in \Fig{fig1}. By contrast, higher $k$
modes, far from the energy cascade inertial range, are notably
inefficient in transferring energy and momentum via charge exchange
mode coupling interactions and are damped by small-scale dissipative
processes in the coupled plasma-neutral heliospheric turbulence.

\section{5. Conclusions}
We have developed a self-consistent fluid model to describe nonlinear
turbulent processes in a partially ionized and magnetized heliospheric
gas. The charge exchange interactions couple the plasma and the
neutral fluids by exciting a characteristic charge exchange coupling
mode $k_{ce}$, which is different from the characteristic turbulent
mode $k$ of the coupled system. One of the most important points to
emerge from our studies is that charge exchange modes modify the
heliospheric turbulence cascades dramatically by enhancing nonlinear
interaction time-scales on large scales.  Thus on scales
$\ell\ge\ell_{ce}$, the coupled plasma system evolves differently than
the uncoupled system where large-scale turbulent fluctuations are
strongly correlated with charge-exchange modes and they efficiently
behave as driven (by charge exchange) energy containing modes of
heliospheric turbulence.  By contrast, small scale turbulent
fluctuations are unaffected by charge exchange modes which evolve like
the uncoupled system as the latter becomes less important near the
larger $k$ part of the turbulent spectrum.  The neutral fluid, under
the action of charge exchange, tends to enhance the cascade rates by
isotropizing the turbulence on a relatively long time scale.  This
tends to modify the characteristics of heliospheric turbulence which
can be significantly different from the Kolmogorov phenomenology of
fully developed turbulence.  We believe that, it is because of this
enhanced nonlinear eddy interaction time, that a large spectral
transfer of turbulent energy tends to smear the current sheets in the
magnetic field fluctuations and further cascade energy to the lower
Fourier modes in the inertial range turbulent spectra. Consequently it
leads to a steeper power spectrum. It is to be noted that the present
model does not consider an external driving mechanism, hence the
turbulence is freely decaying.  Driven turbulence, such as due to
large scale external forcing, e.g. supernova explosion, may force
turbulence at larger scales. This can modify the cascade dynamics in a
manner usually described by dual cascade process.


The support of NASA(NNG-05GH38) and NSF (ATM-0317509) grants is  acknowledged.





\IfFileExists{\jobname.bbl}{}
 {\typeout{}
  \typeout{******************************************}
  \typeout{** Please run "bibtex \jobname" to optain}
  \typeout{** the bibliography and then re-run LaTeX}
  \typeout{** twice to fix the references!}
  \typeout{******************************************}
  \typeout{}
 }


\end{document}